\documentclass[journal]{IEEEtran}

\pdfoutput=1
\usepackage{epsfig,floatflt}
\usepackage{graphicx,cite,url}
\usepackage{subfigure}
\usepackage{amsmath,amsfonts,amssymb}
\usepackage{algorithm,algorithmic}

\IEEEoverridecommandlockouts

\nonstopmode

\begin{document}

\title{PHY-layer link quality indicators for wireless networks using
matched-filters}

\author{Henry E. Baidoo-Williams\IEEEauthorrefmark{1}, Octav Chipara\IEEEauthorrefmark{1}, \IEEEauthorblockN{Raghuraman
Mudumbai\IEEEauthorrefmark{1} and Soura Dasgupta\IEEEauthorrefmark{1}\\}
\authorblockA{\IEEEauthorrefmark{1}The University of Iowa, Iowa City, IA 52242,
USA\\} }

\maketitle

\begin{abstract}

We present a novel approach to accurate real-time estimation of wireless link
quality using simple matched-filtering techniques. Our approach is based on the
simple observation that there is a portion of each packet transmission from any
given node that does not change from one packet to another; this includes
preamble sequences used to synchronize the receiver and also address
information in the packet header used for medium access control and routing.
Our approach can be thought of as a generalized and simplified variant of
standard signal processing techniques that are commonly used for preamble
detection, automatic gain control, carrier sensing and other functions in many
packet wireless networks. By using a combination of energy detection and
correlation techniques, we show that we can effectively detect packet
transmissions in real-time with low complexity, without decoding the packets
themselves, and indeed, even without detailed knowledge of the packet format.
We present extensive experimental results from a software-defined radio testbed
to illustrate the effectiveness of this approach for 802.15.4 (Zigbee) networks
even in the presence of strong interference signals and low SNR.
 
\end{abstract}

\section{Introduction} \label{sec:intro}

This paper is motivated by the problem of link quality estimation in wireless
networks. Previous work has established that availability of link quality
indicators (LQI) can substantially improve the performance of medium access
control (MAC) and routing protocols in wireless networks. Consequently, there
has been a lot of work on designing networking protocols that assume accurate
and timely LQI estimates. Furthermore the accuracy and timeliness of the LQI
information has a significant effect on network performance. However, existing
LQI estimators are very crude, being constrained by very minimal hardware
support and protocol layering considerations. We present in this paper a novel
approach using matched-filtering techniques that can provide significantly
richer LQI information with minimal overhead and processing complexity.

\subsection{Motivation and problem statement}

The main challenge in obtaining good estimates is the overhead cost in terms of
power consumption and complexity. Wireless link states vary rapidly because of
fading and mobility effects, so LQI estimates need to be frequently updated.
Receiver hardware is usually powered off unless a packet is detected for
decoding, so there is typically very limited hardware support for continuous
link monitoring. Usually the hardware support takes the form of a quantized
received signal strength information (RSSI) signal that is updated frequently
and continuously (typically every few symbol intervals). While the RSSI signal
can be used to flag activity or inactivity in the medium, it cannot distinguish
between different transmitters or interferers, and as a result, the RSSI signal
cannot by itself be used to track the states of multiple links.

It is possible to augment the RSSI signal with other information to obtain
better LQI estimates. For instance, apriori knowledge of transmission schedules
of different nodes and channel fading statistics can be combined with RSSI to
track link states. Packet failure rates and large link latencies have also been
proposed as proxies to indicate a weak link. However, such additional
information is not always available and such techniques often depend on strong
modeling assumptions (e.g. long channel coherence times).

It is also possible to combine RSSI information with knowledge from the packet
decoder about successful packet receptions for LQI estimation. However, this
only works for successfully decoded packets, and does not provide information
about transmissions not intended for the specific receiver. While, in
principle, it is possible to configure the packet decoder to listen
promiscuously to all transmissions in the network, in practice this has two
serious limitations: (a) this greatly increases power consumption, and (b) this
does not provide any information about packets that fail to decode because of
collisions, interference or fading. This is an especially serious limitation on
networks operating in a shared part of the frequency spectrum e.g. the ISM band
or a cognitive radio application, where the interfering signal may come from a
transmission on another network using completely different protocols and packet
formats. Finally, there is limited standards support for cross-layer
information sharing to facilitate more sophisticated LQI estimation.

\subsection{Summary of contributions}

We propose a novel approach for accurate, real-time LQI estimation using
matched-filters and present an extensive set of experimental results that
demonstrate the effectiveness of our approach. The basic idea behind our
approach is to exploit the fact that there is a significant amount of
redundancy in transmissions in modern wireless networks: every packet sent
between the same pair of nodes have a significant number of symbols that are
identical across packets. Thus if we correlate the received signal with a known
sequence of symbols using a matched-filter, a sharp peak in the filter output
is a good indicator of the presence of the symbols. The size of the peak
provides an estimate of the link channel strength.

Additionally, there are a number of symbols (e.g. source and destination
addresses) that are always distinct for packets between different pairs of
nodes. Thus we can use a bank of parallel filters each matched to a different
set of symbols to identify specific transmitters.

Our main contributions are summarized as follows.
\begin{enumerate}

\item We describe an algorithm using peak detection on the output of
matched-filtering complex baseband samples of the signal at a receiver to
accurately detect, identify and classify packet transmissions.

\item We show that our matched filtering approach can effectively detect and
identify packet transmissions even without any knowledge of the packet format,
simply by using a noisy recorded copy of a previous transmission as a template
for the matched-filter.

\item While continuously-running matched filters can be expensive in terms of
computation and power consumption, we show that a simple energy detector with a
threshold to trigger the filters on when an incoming transmission is detected,
works well to minimize the overhead cost of the filters.

\end{enumerate}

We implemented these ideas on a testbed using Zigbee transmitters and a
software-defined radio receiver and verified their effectiveness for LQI
estimation.

\subsection{Background and related work}

Network and link-layer protocols for wireless network protocols share a common
lineage with those for wired networks; however it is well-known that wireless
networks has peculiarly challenging features such as a large range of variation
of signal power \cite{dynrange}, fading and time-varying channels
\cite{timevarying} and link asymmetries \cite{asymmetries}, hidden and exposed
terminals \cite{hidden_term} and so on. Link quality estimation has long been
recognized as a classical problem for wireless networks \cite{lqi_1990} as a
way of addressing these challenges. More recently the development of ad-hoc and
sensor networks has led to a renewed surge of interest in this topic
\cite{couto2005high}. Several protocols have been designed that can take
advantage of LQI information to realize significant improvements in network
performance \cite{prot1, prot2, prot3}, and the IEEE 802.15.4 standard includes
specifications for LQI \cite{zigbee_lqi}.

However the problem of LQI estimation still remains very much open
\cite{indoor_lqi_survey, lqi_multihop}. One important reason for this is that
protocol layering constraints have limited effective sharing of information
relevant to LQI estimation from the PHY and link layers to higher level
protocols \cite{fourbit_lqi}. There is also a tradeoff between the overhead
cost of continuously monitoring transmissions on the medium and the accuracy of
the measured LQI information. Partly because of these tradeoffs, many LQI
estimation techniques \cite{lqi_wsn} rely on packet success probabilities and
similar measures that can be easily monitored by higher layer protocols.
Hardware support for LQI is most commonly available in the form of a quantized
received signal strength indicator (RSSI); RSSI signals are useful not just for
LQI, but also for performing energy detection to assist in carrier-sensing
\cite{cca} and other MAC functions. A comparative study of the performance of
commonly used LQI estimation metrics is presented in \cite{lqi_survey1}.

\subsubsection{Relationship to preamble-detection filters}

All packet networks use a preamble sequence \cite{preamble} at the beginning of
packet transmissions to aid the receiver in detecting and ``acquiring" the
incoming signal \cite{frame_sync}, a process that typically involves
determining the frame boundary, symbol boundaries and carrier frequency and
phase offsets. The preamble sequences are carefully designed to have a sharply
peaked auto-correlation function, and matched-filters are universally used to
detect the preambles and estimate their precise timing. The preamble-detection
filters are also sometimes used for coherent carrier-sensing \cite{cca} for
medium access control.

Our approach is also based on this same idea of detecting correlation peaks; so
a natural question is: can we leverage the preamble-detection filter to also
perform LQI estimation perhaps with some minor modification? This is an
attractive possibility because it could potentially give us LQI estimation
free-of-cost simply as a by-product of the signal acquisition process.

However, the LQI estimation process differs in some fundamental respects from
preamble-detection. Specifically, the preamble detection function is expected
to produce a very fine-grained estimate of the timing of the correlation peak,
because the carrier and symbol timing synchronization algorithms depend
sensitively on this estimate. Also preamble detection errors can be extremely
costly in terms of performance, e.g. false alarms trigger spurious collision
detects which can degrade the medium-access control function while missed
detection events lead to packet losses.

Thus, preamble detection filters are typically implemented with very high
precision computations e.g. with 16-bit quantized samples. This can be
extremely power-hungry, and indeed there has been some interesting recent work
\cite{preamble_detection} on dynamically switching between high and low
precision calculations to minimize the power consumption.

Since LQI estimation is far more tolerant of estimation errors, it is not
obvious what is the optimal way to jointly implement this function with other
PHY tasks. For the purposes of this paper, we assume that the different matched
filters used at the receiver are all independent and running in parallel and we
defer the details of their optimal implementation for future work.

\subsection{Outline}

The rest of the paper is organized as follows. In Section \ref{sec:mf} we
describe our proposed technique for link estimation. We present a detailed set
of experimental results demonstrating the performance of our approach in
Section \ref{sec:results}. Specifically, we consider the ``protocol-aware" case
in Section \ref{sec:protaware} (where the receiver knows the packet format and
has a noiseless copy of a known sequence of samples for each link being
monitored), and Section \ref{sec:protblind} addresses the ``protocol-blind"
case (where the links being monitored use packet formats unknown to the
receiver). Section \ref{sec:conc} concludes.

\section{Proposed technique for signal detection using matched-filters}
\label{sec:mf}

We describe our proposed technique for signal detection using matched filters. The goal of the detection procedure is simply to tell if a packet was transmitted and if so from which transmitter. The receiver node, Rx0, which does the detection is depicted in Fig. \ref{fig:detector-node} below. In our setup, Rx0 is an ettus USRP N200 with a RFX2400 daughter board. 

\begin{figure}[htb]
\begin{center}
	\includegraphics[width=3in]{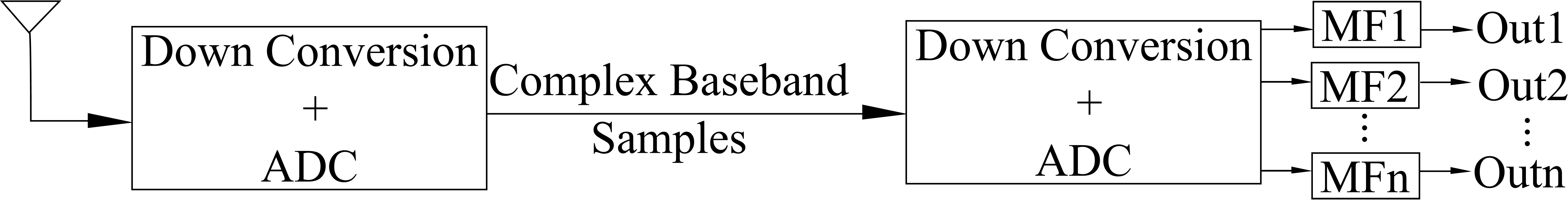} 
\caption{Signal detection scheme}
\label{fig:detector-node}
\end{center}
\end{figure}

The MF block as shown in Fig. \ref{fig:detector-node} is the block that performs all the relevant discrimination of received samples. There are three major steps in this block which are controlled by three threshold applications. The first step is to detect whether the received signal is above the noise floor. An energy detection threshold T1 is applied. If the energy of the received signal is above T1, then the matched filter value is computed else it is set to zero. This ensures that energy is conserved until there is a potential packet arriving. The matched filter output denoted $mf$, given N filter taps $h$ and received signal samples, $x$ is found by
\begin{align}
{mf[i]} = \sum_{k=i-N+1}^i h[i-k]^*x[k] \label{eq:mf1}
\end{align}

The second step is to detect whether there is a relevant match between the received signal and the filter taps after T1 has been satisfied. This is checked using a simple metric, denoted $m$, given by
\begin{align}
{m[i]} = \frac{mf[i]}{mf[i-1]} > {T2} \label{eq:mf2}
\end{align}
It should be noted that when there is a packet, the value of the matched filter output is bell shaped. Hence there has to be a metric value greater than 1 if there is a packet matching the taps present in the captured samples. To reduce the number of computations, values of the matched filter threshold T2 greater than 1 can be utilised.

The third step is to evaluate the correlation coefficient given that thresholds T1 and T2 have been satisfied. We use the Pearson correlation coefficient, denoted $c$, thus
\begin{align}
{c[i]} = \frac{mf[i] - \frac{\displaystyle\sum\limits_{k=1}^N h[k] \displaystyle\sum\limits_{k=z}^i x[k]}{N}}{\sqrt{(\displaystyle\sum\limits_{k=z}^i x^2[k]-\frac{(\displaystyle\sum\limits_{k=z}^i x[k])^2}{N})(\displaystyle\sum\limits_{k=1}^N h^2[k]-\frac{(\displaystyle\sum\limits_{k=1}^N h[k])^2}{N})}} \label{eq:mf3}
\\*{z} = {i-N+1}
\end{align}
The third threshold T3 is finally applied to the output of $c$. If the output meets the threshold it is kept, else the value is set to zero. The norm of $c$ is between 0-1 and a value close to 1 denotes a match. Satisifying all three thresholds implies there was a packet related to a particular match filter and its filter taps.

The pseudo-code for this algorithm described above is given in algorithm \ref{alg:sigdetect}.

\begin{algorithm}
\caption{Signal detection algorithm at each MF block}
\label{alg:sigdetect}

\begin{algorithmic}
\STATE \textbf {Initialization:}
\STATE $T1 \leftarrow received\_signal\_strength\_threshold$
\STATE $T2 \leftarrow matched\_filter\_metric\_threshold$
\STATE $T3 \leftarrow correlation\_coefficient\_threshold$
\WHILE {$receiver\_Rx0\_is\_running$}
\IF {$received\_signal\_strength > T1$} 
        \STATE Compute $matched\_filter\_output$
        \STATE Compute $matched\_filter\_metric\_output$
        \IF {$matched\_filter\_metric\_output > T2$}
                \STATE Compute $correlation\_coefficient\_output$\
                \IF {$correlation\_coefficient\_output <= T3$}
                \STATE $correlation\_coefficient\_output \leftarrow 0$
                \ENDIF
        \ELSE
                \STATE  $correlation\_coefficient\_output \leftarrow 0$
        \ENDIF
\ELSE
        \STATE $matched\_filter\_output \leftarrow 0$
        \STATE $correlation\_coefficient\_output \leftarrow 0$
\ENDIF 
\ENDWHILE

\end{algorithmic}
\end{algorithm}

\section{Experimental results} \label{sec:results}

\begin{figure}[htb]
\begin{center}
	\includegraphics[width=3in]{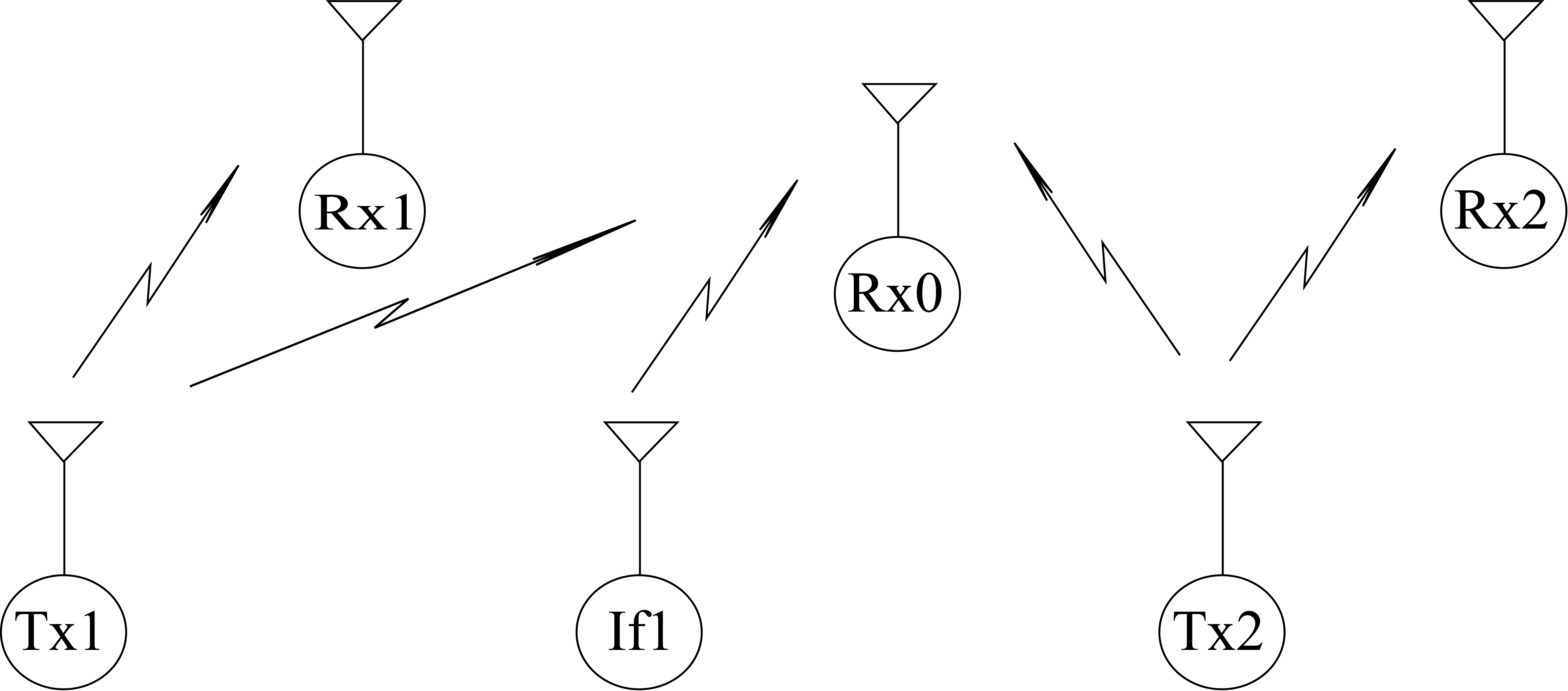} 
\caption{Experiment setup for signal detection.}
\label{fig:exp-setup}
\end{center}
\end{figure}

We now present experimental results from our implementation. Fig. \ref{fig:exp-setup} shows a diagram of our experimental setup which comprises two transmitter nodes, Tx1 and Tx2 with corresponding receivers Rx1 and Rx2, an interfering signal transmitter, If1, and an eavedropper Rx0 within range of both Tx1 and Tx2 which is going to be used to detect the presence or absence of packets transmitted by Tx1 and Tx2. The hardware equipment used for the two types of experiments, thus protocol-aware signal detection and protocol-blind signal detection are described in the sections below with the observed results.

\subsection{Protocol-aware signal detection} \label{sec:protaware}
\begin{figure}[htb]
\begin{center}
	\includegraphics[width=3in]{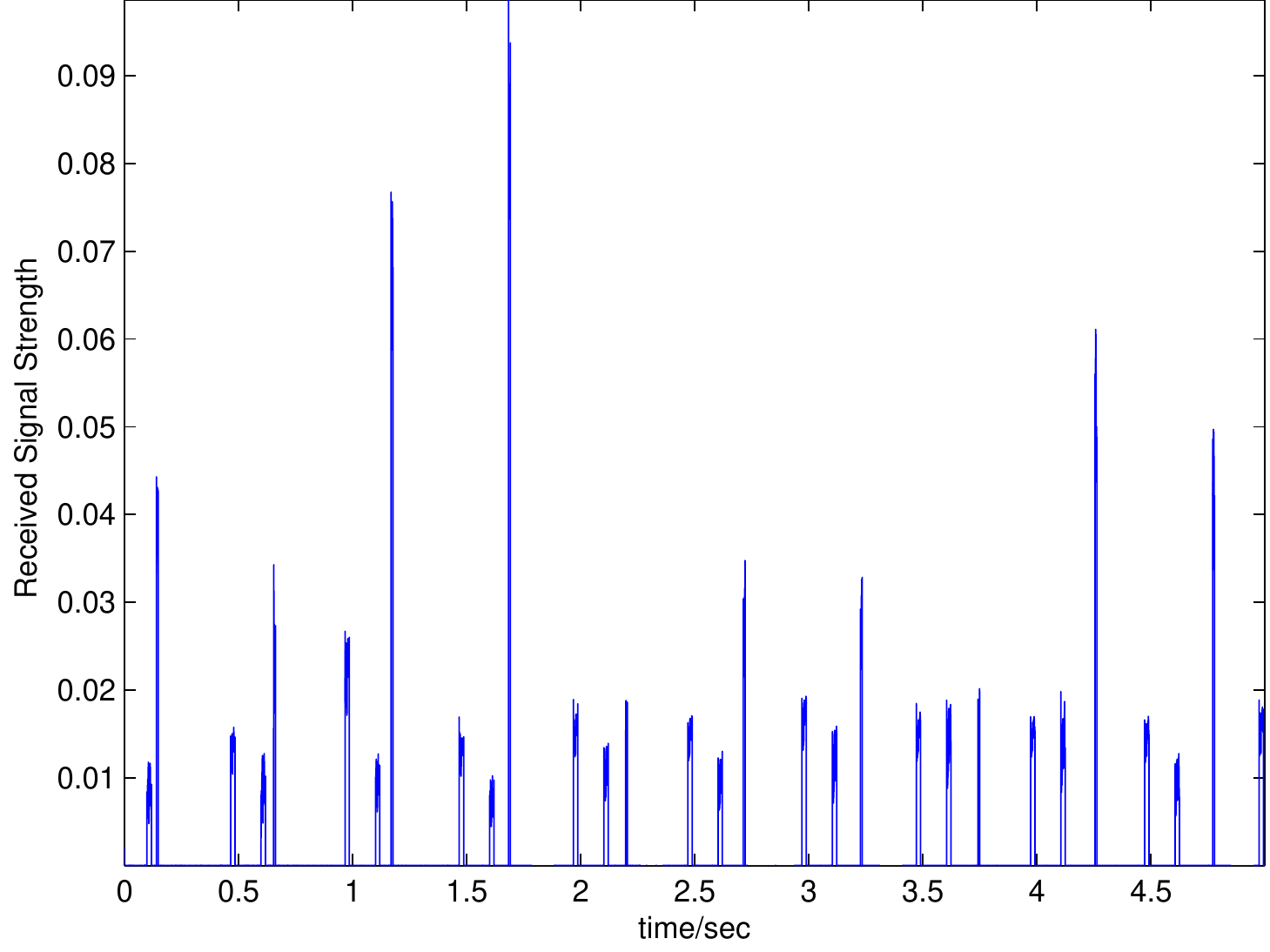} 
\caption{Received RSS at Rx0 using USRP N200 transmitters}
\label{fig:rssusrp}
\end{center}
\end{figure}

\begin{figure}[htb]
\begin{center}
	\includegraphics[width=3in]{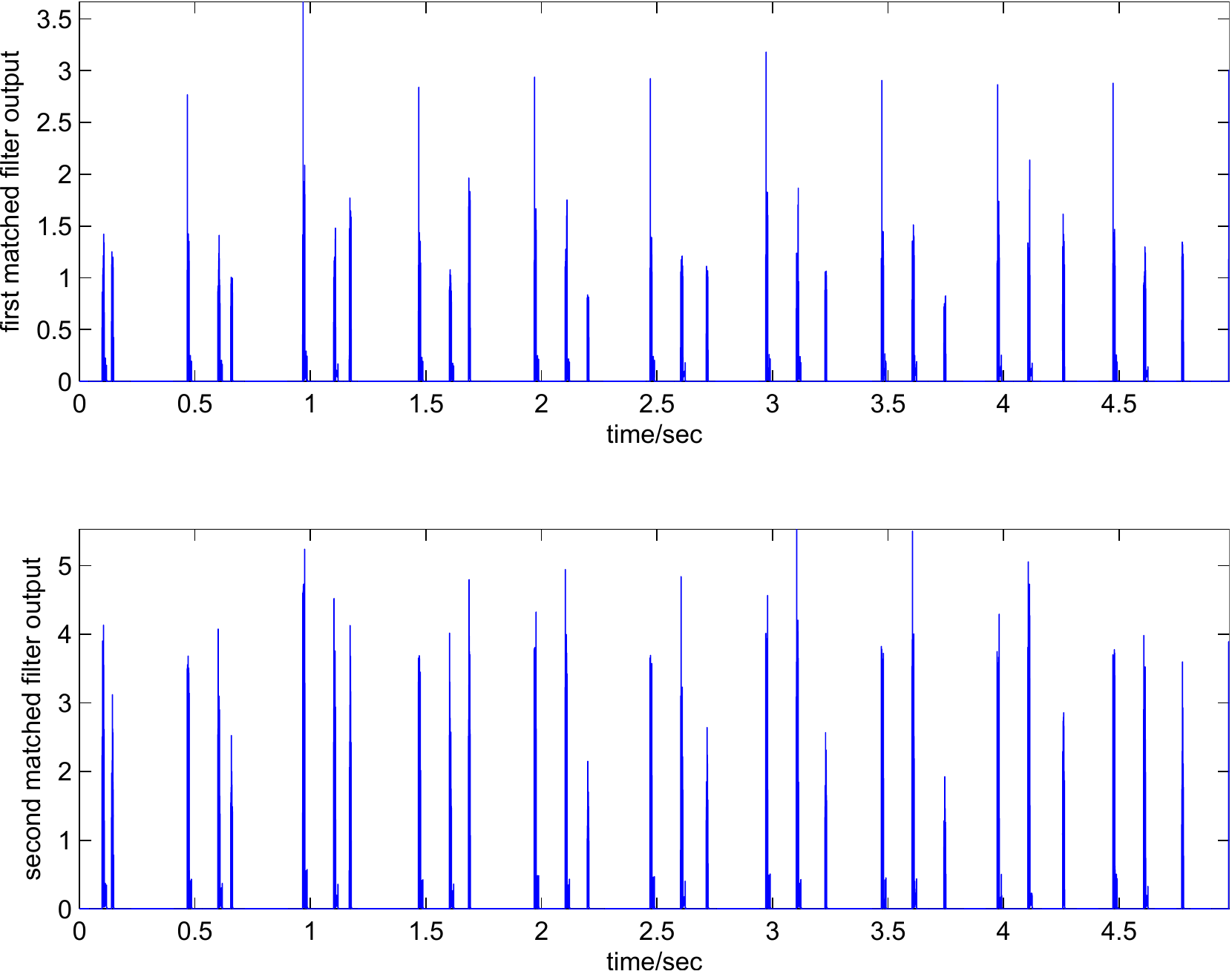} 
\caption{Matched filter output at Rx0 using USRP N200 transmitters}
\label{fig:mfusrp}
\end{center}
\end{figure}

\begin{figure}[htb]
\begin{center}
	\includegraphics[width=3in]{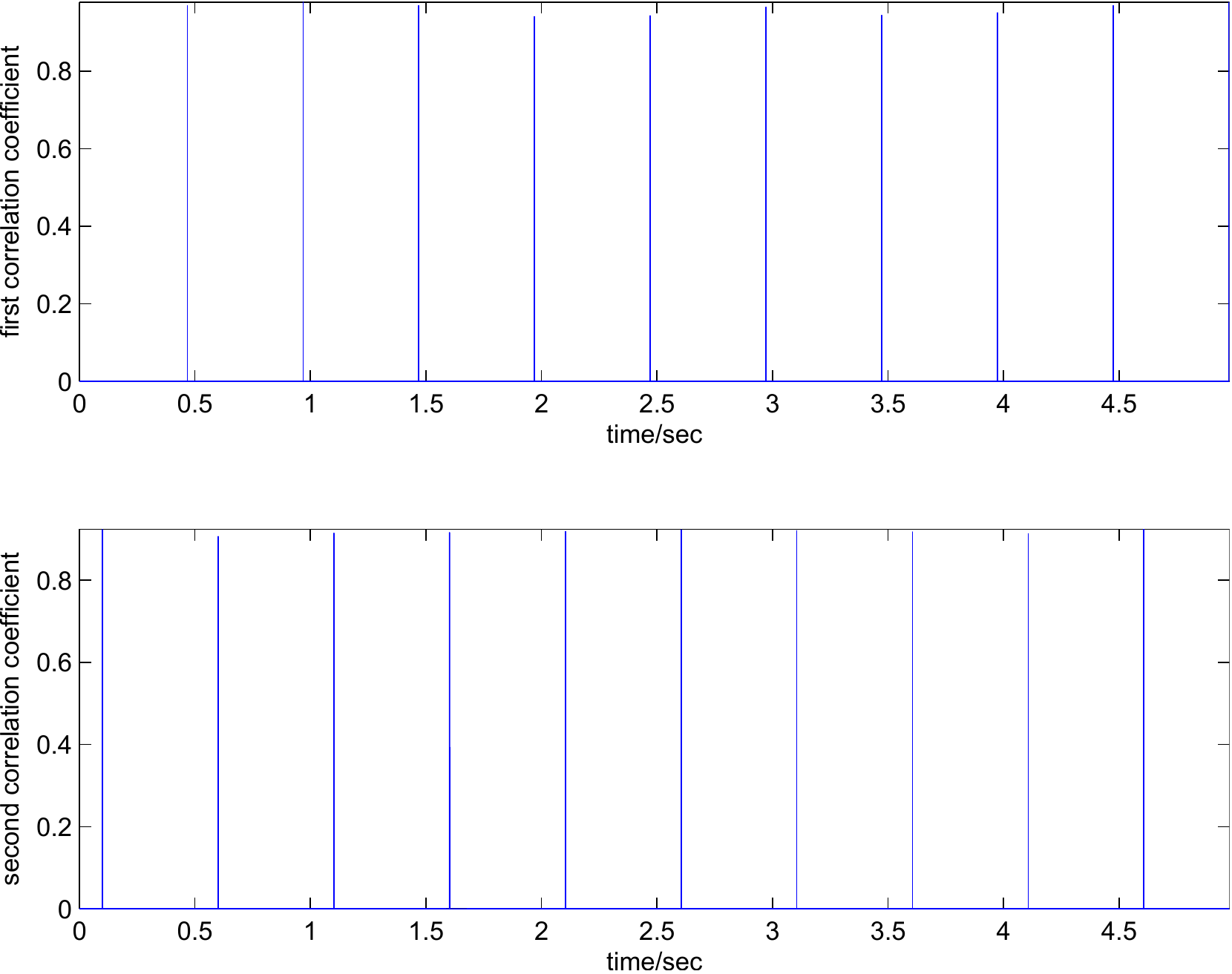} 
\caption{Correlation coefficient output at Rx0 using USRP N200 transmitters}
\label{fig:corcoeffusrp}
\end{center}
\end{figure}

In the protocol-aware signal detection setup, Tx1, Tx2, If1, Rx0, Rx1 and Rx2 are all USRP N200 nodes with RFX2400 daughter boards. The pre-recorded packet samples used for the two filter taps corresponding to Tx1 and Tx2 are 128 samples generated from the one byte start frame delimiter (SFD). Two different SFD values of 0xA7 and 0x98 are used to distinguish between the two packets. The UCLA zigbee PHY package is used for generating the zigbee packets \cite{schmid2006gnu} \cite{choong2009multi}. The USRP N200 node If1 transmits pre-recorded gmsk packets as an interfering signal. All signals are transmitted at a Radio frequency(RF)  of 2.48GHz. The parameters used for the experiment at Rx0 are energy detection threshold T1=0.01, matched filter metric threshold T2=1 and correlation coefficient threshold T3=0.8 for both filters. Tx1, Tx2 and If1 all transmit one packet every half second. Results of the received signal strength(RSS), matched filter output and correlation coefficient output are shown in  Figs. \ref{fig:rssusrp}-\ref{fig:corcoeffusrp}.

From Figs. \ref{fig:rssusrp}-\ref{fig:corcoeffusrp}, it is clear that the RSS plot shows all three signals, thus the two desired packets as well as the interfering signal, whiles the matched filter plot also looks similar to RSS plot albeit at different scales. The correlation coefficient plot however separates the two desired packets. The correlation coefficient plot corresponding to the packets from Tx1 completely eliminates the packets from Tx2 and If1. A similar observation holds for the correlation coefficient plot corresponding to packets from Tx2.

\subsection{Protocol-blind signal detection} \label{sec:protblind}
\begin{figure}[htb]
\begin{center}
	\includegraphics[width=3in]{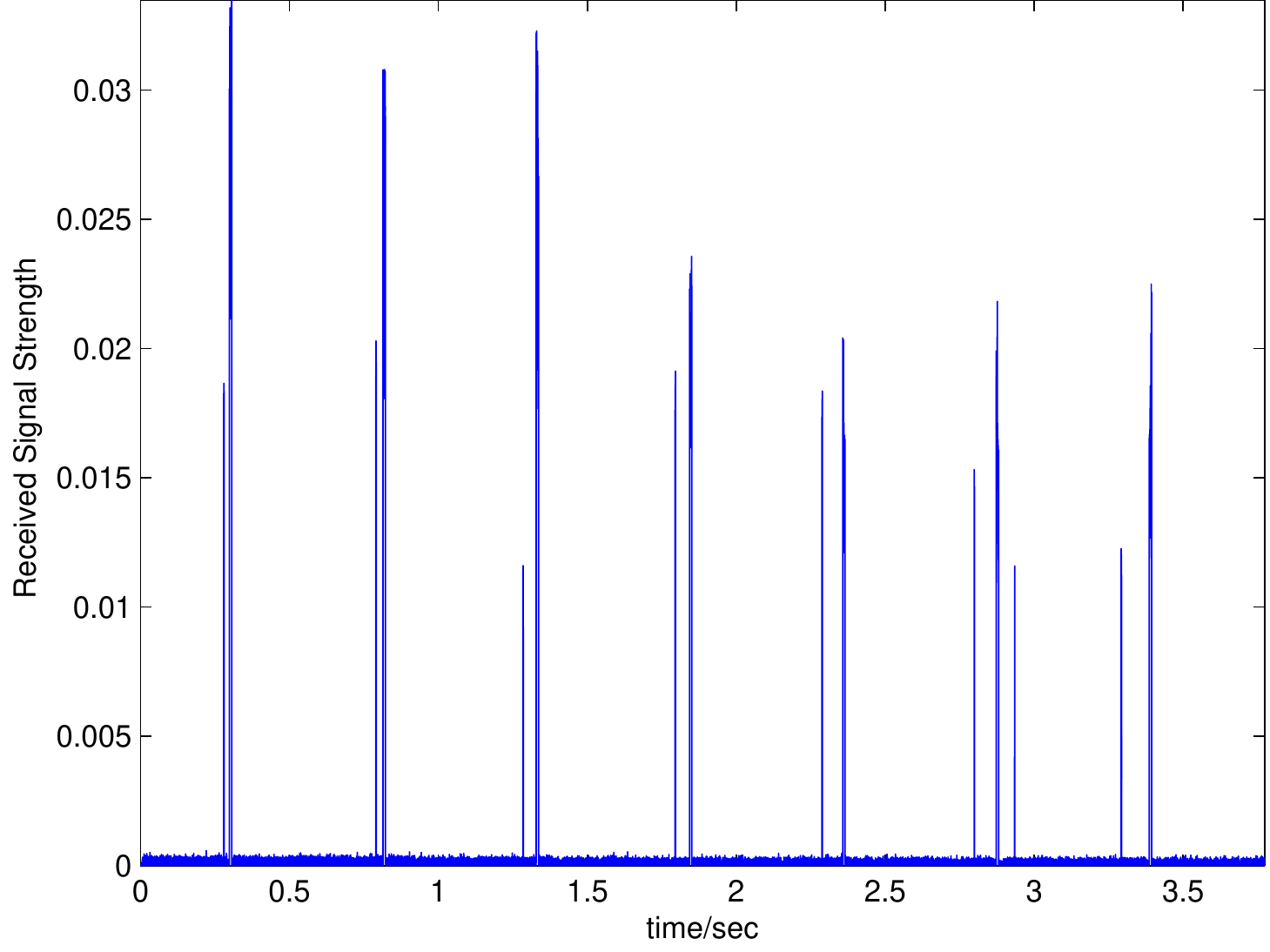} 
\caption{Received RSS at Rx0 using telosb motes as transmitters}
\label{fig:rsstelosb}
\end{center}
\end{figure}

\begin{figure}[htb]
\begin{center}
	\includegraphics[width=3in]{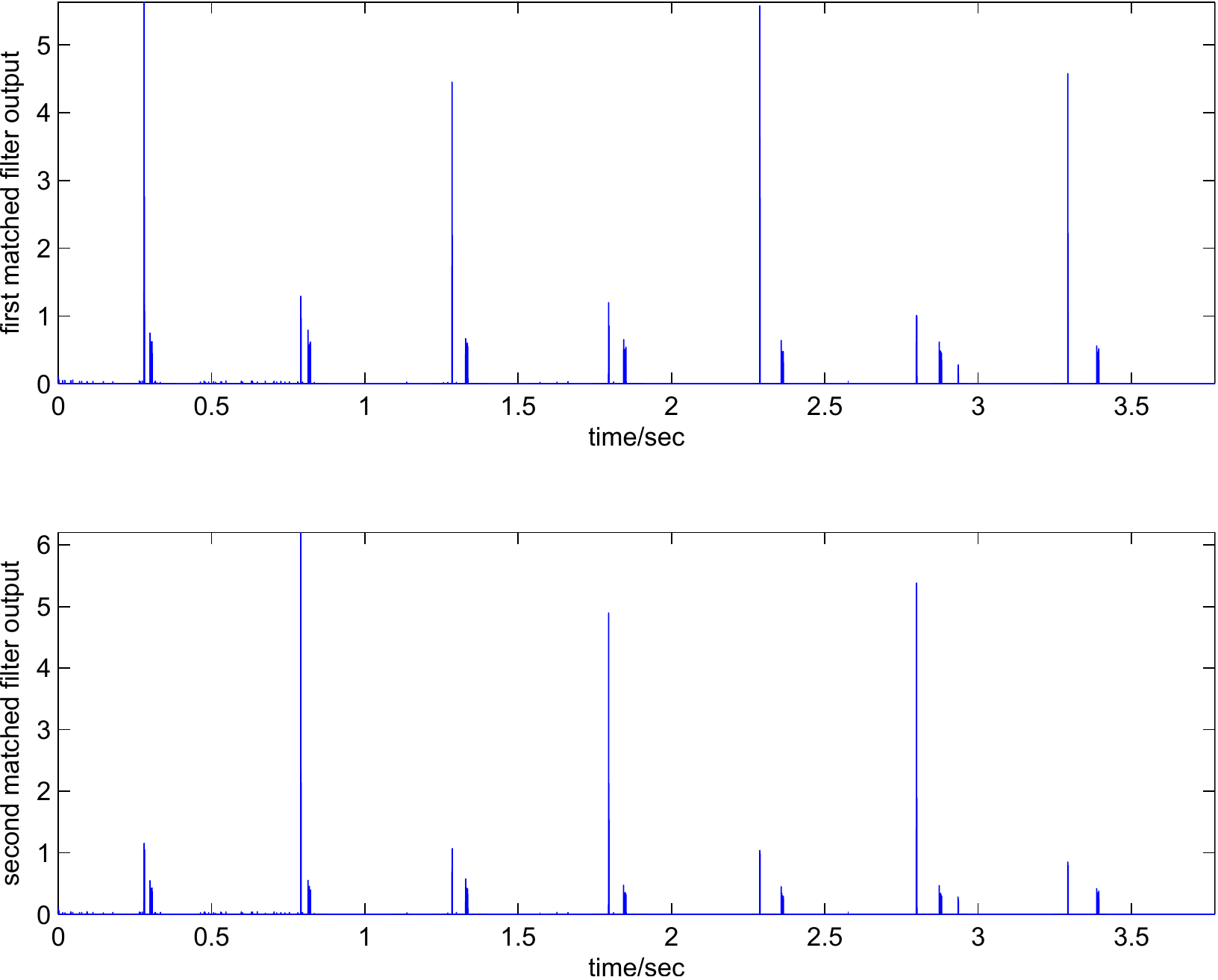} 
\caption{Matched filter output at Rx0 using telosb motes as transmitters}
\label{fig:mftelosb}
\end{center}
\end{figure}

\begin{figure}[htb]
\begin{center}
	\includegraphics[width=3in]{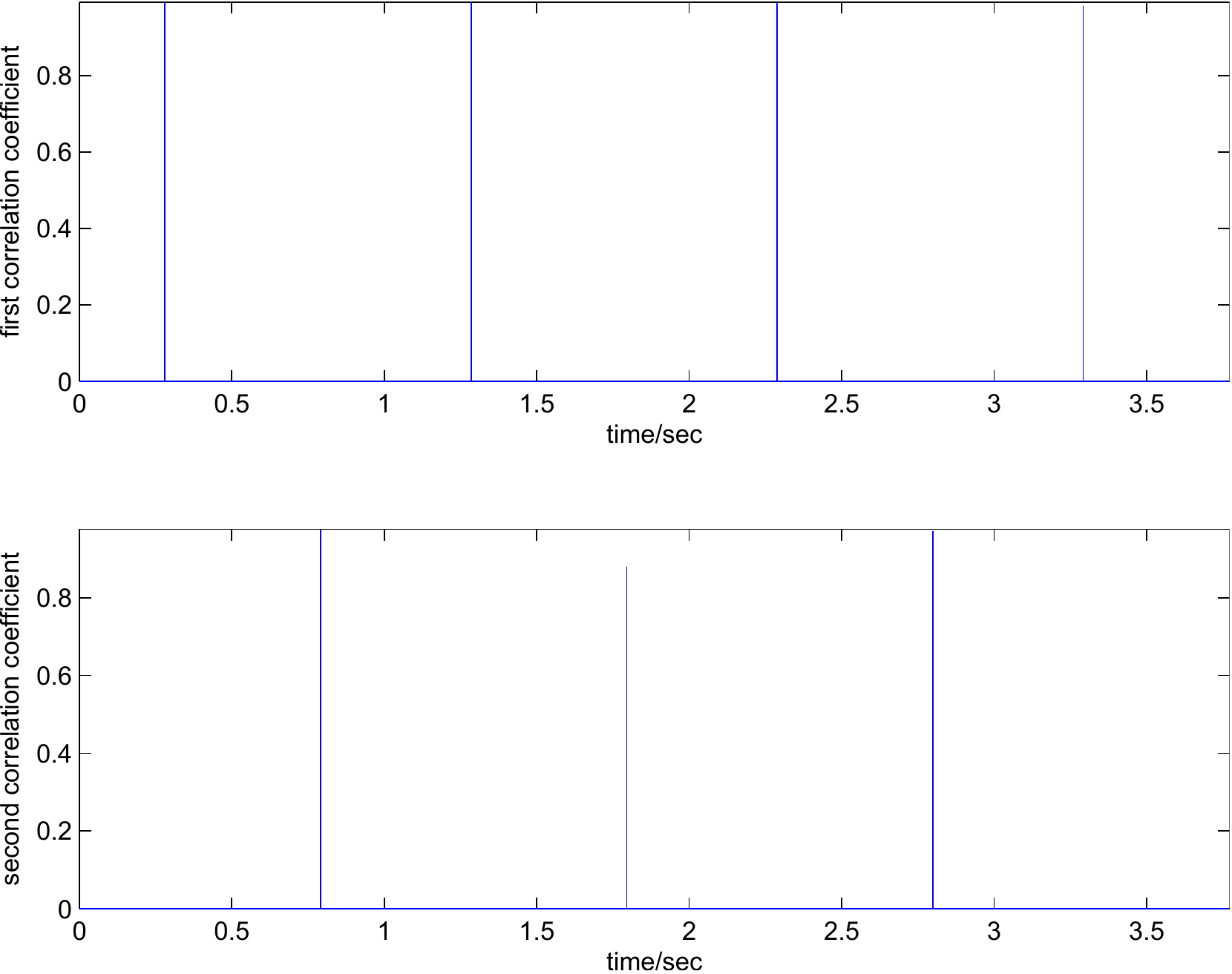} 
\caption{Correlation coefficient output at Rx0 using telosb motes as transmitters}
\label{fig:corcoefftelosb}
\end{center}
\end{figure}

In the protocol-blind signal detection setup, Tx1, Tx2, Rx1 and Rx2 are crossbow telosb motes whiles If1 and Rx0 are USRP N200 nodes. The pre-recorded packets samples used for the two filter taps corresponding to Tx1 and Tx2 packets are 576 samples generated from four bytes in the address information. Specifically the AM broadcast address information is used. The samples however are noisy and are recorded at the Rx0 at a sampling rate of 4MHz. Two different AM broadcast address information of 0xAAAA and 0x37BD are used for the two motes to distinguish between the two packets. Tinyos 2.0.2 is used for generating the packets and compiling the binary image unto the telosb motes. The USRP N200 node If1 again transmits pre-recorded gmsk packets as an interfering signal. The parameters used for the experiment at Rx0 are energy detection threshold T1=0.01, matched filter metric threshold T2=1 and correlation coefficient threshold T3=0.8 for both filters. Tx1 and Tx2 transmit one packet every second whiles If1 transmits a packet every half second. Results of the RSS, matched filter output and correlation coefficient output are shown in  Figs. \ref{fig:rsstelosb}-\ref{fig:corcoefftelosb}.

From Figs. \ref{fig:rsstelosb}-\ref{fig:corcoefftelosb}, it is clear that the RSS plot shows the two desired packets as well as the interfering signal, whiles the matched filter plot also looks similar to RSS plot but has greater values for the desired packets. The correlation coefficient plot however clearly separates the two desired packets even though the filter taps are noisy. The correlation coefficient plot corresponding to the packets from Tx1 completely eliminates the packets from Tx2 and If1. A similar observation holds for the correlation coefficient plot corresponding to packets from Tx2.

\begin{figure}[htb]
\begin{center}
	\includegraphics[width=3in]{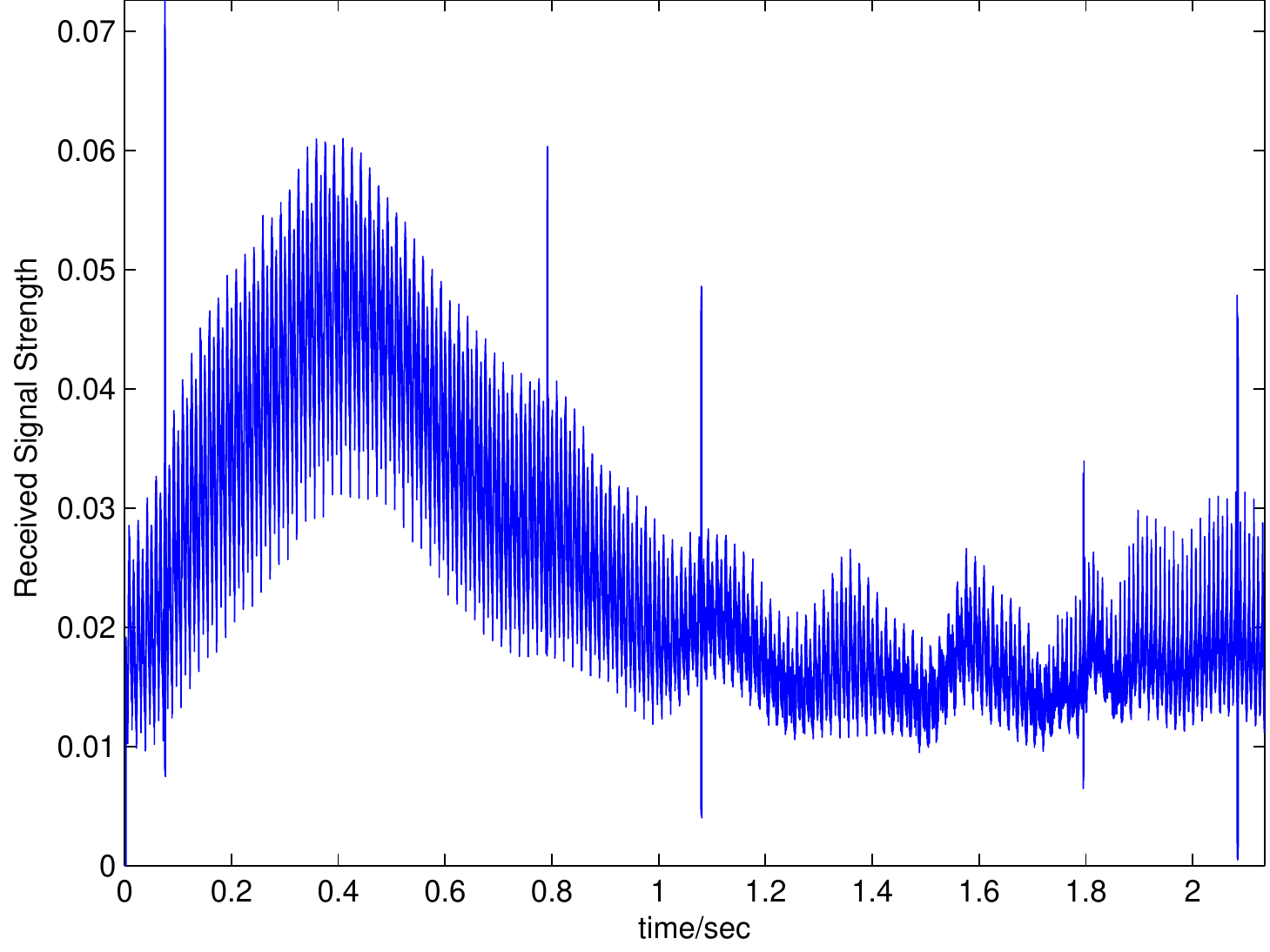} 
\caption{Received RSS at Rx0 using telosb motes as transmitters}
\label{fig:rsstelosb2}
\end{center}
\end{figure}

\begin{figure}[htb]
\begin{center}
	\includegraphics[width=3in]{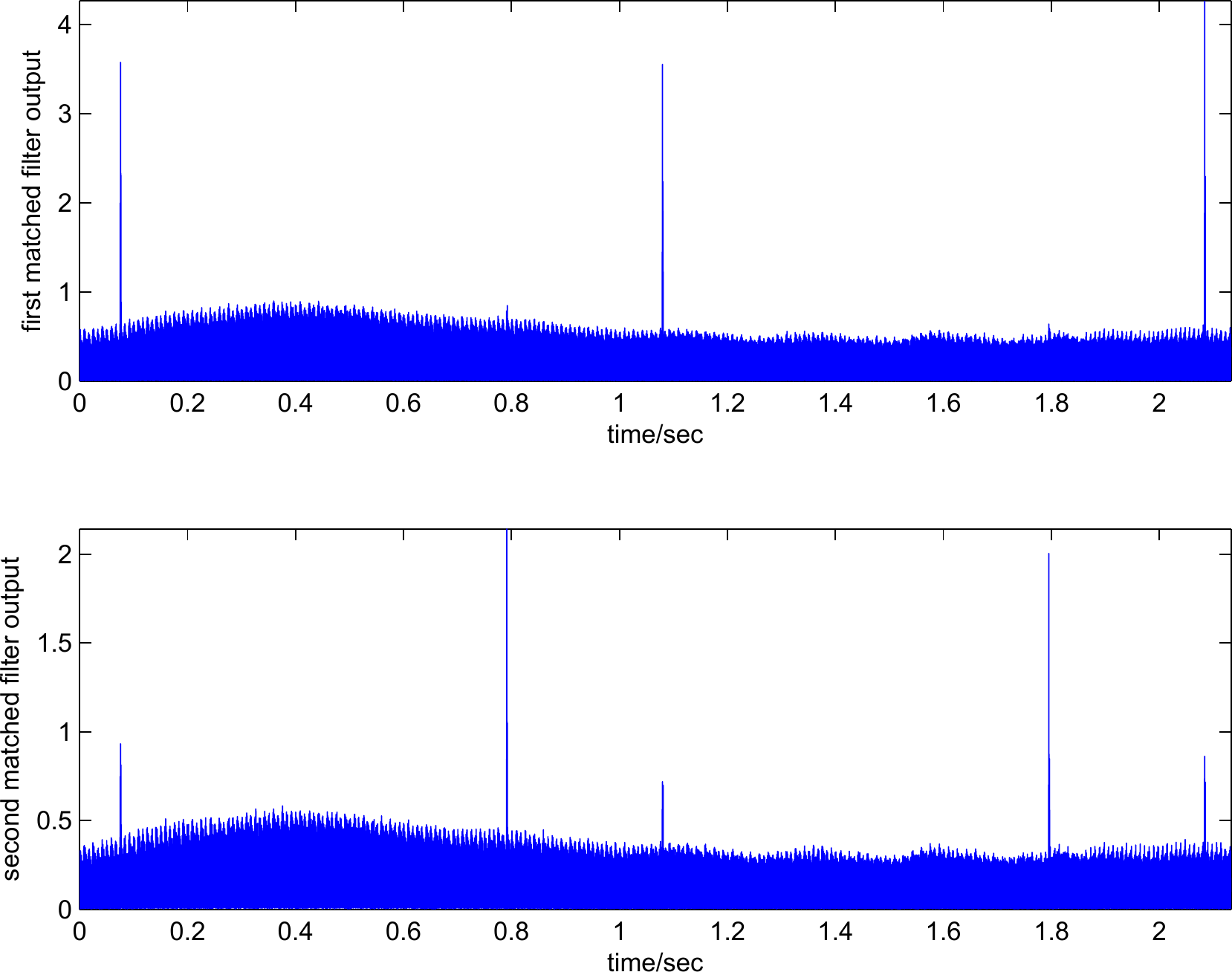} 
\caption{Matched filter output at Rx0 using telosb motes as transmitters}
\label{fig:mftelosb2}
\end{center}
\end{figure}

\begin{figure}[htb]
\begin{center}
	\includegraphics[width=3in]{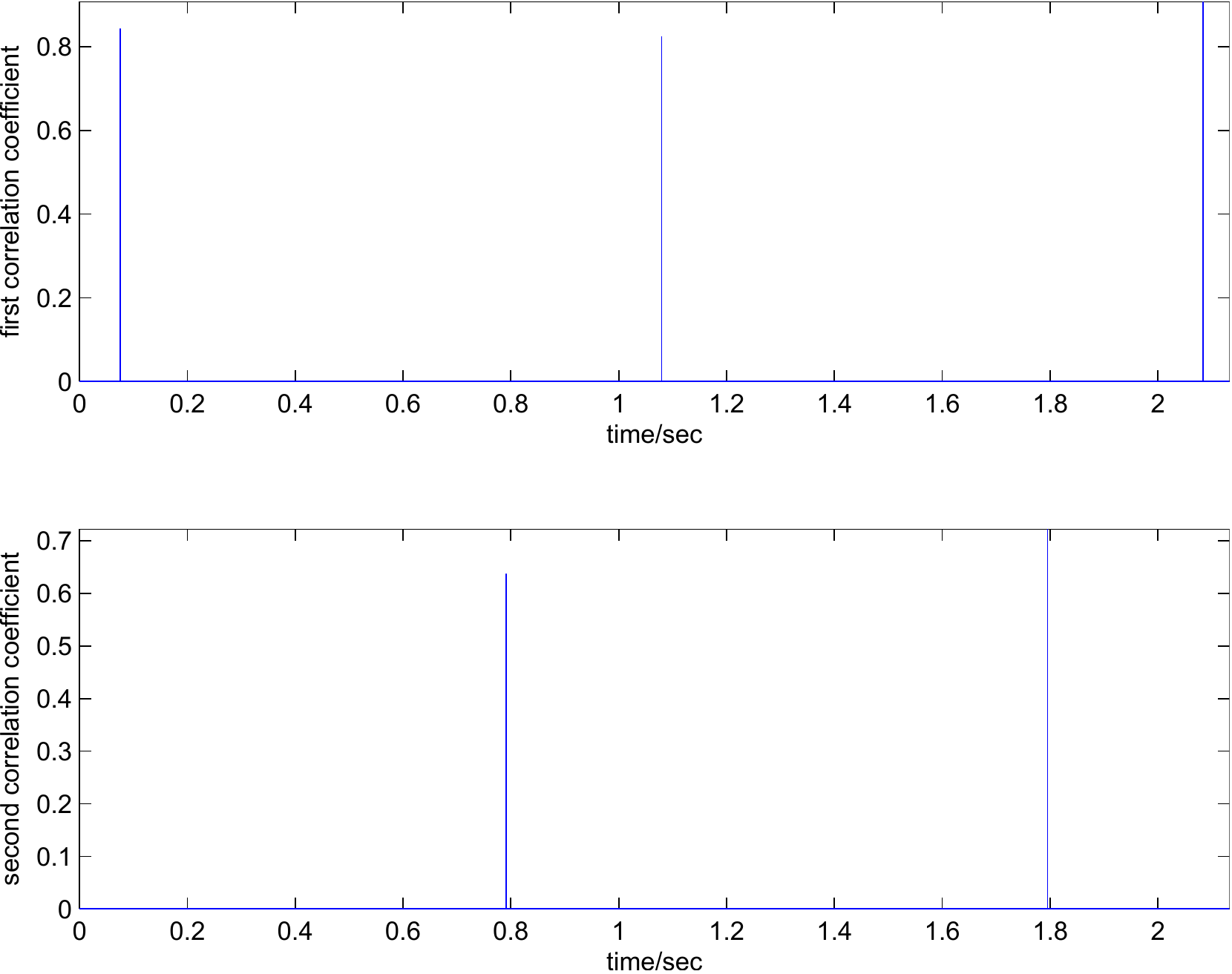} 
\caption{Correlation coefficient output at Rx0 using telosb motes as transmitters}
\label{fig:corcoefftelosb2}
\end{center}
\end{figure}

Another experiment was conducted to clearly show that even in the case where the interfering signal coincides with desired packets, we are still able to detect the packets. In this experiment, the parameters for Tx1 and Tx2 are maintained as above but If1 now transmits a constant continuous burst at the same RF frequency of the two telosb motes.  The parameters used for the experiment at Rx0 are energy detection threshold T1=0.01, matched filter metric threshold T2=1 and correlation coefficient threshold T3=0.5 for both filters. It is worthy of note that reducing the signal-to-noise-plus-interference (SNIR) only affects the maximum value of the peak correlation coefficient and therefore necessitates reducing the correlation coefficient threshold appropriately.  Results of the received signal strenghth(RSS), matched filter output and correlation coefficient output are shown in  Figs. \ref{fig:rsstelosb2}-\ref{fig:corcoefftelosb2}.

From Figs. \ref{fig:rsstelosb2}-\ref{fig:corcoefftelosb2}, it is seen that the two packets are again easily distinguishable even at SNIR below 0dB since the transmit power of the telosb motes were kept at the same level as the previous experiment. Reducing the SNIR only reduces the maximum peak of the correlation coefficient value. The correlation coefficient threshold will therefore have to be sized appropriately.

\section{Conclusion} \label{sec:conc}

\bibliographystyle{IEEEtran}
\bibliography{refs}

\end{document}